\documentclass[a4paper]{article}

\catcode`\é=13  \defé{\'e}
\catcode`\â=13  \defâ{\^a}
\catcode`\è=13  \defè{\`e}
\catcode`\à=13  \defà{\`a}
\catcode`\ù=13  \defù{\`u}
\catcode`\ç=13  \defç{\c c}
\catcode`\ê=13  \defê{\^e}
\catcode`\ô=13  \defô{\^o}
\catcode`\û=13  \defû{\^u}
\catcode`\ù=13  \defù{\`u}
\catcode`\î=13  \defî{\^\i}
\catcode`\ï=13  \defï{\"\i}

\usepackage{fullpage} 
\usepackage{pstricks}
\usepackage{amssymb}
\usepackage{amsthm}
\usepackage{amsmath} 
\usepackage{amsfonts}
\usepackage{epstopdf}
\usepackage[version=3]{mhchem}
\usepackage{multicol}
\usepackage{hyperref}
\usepackage{graphicx}
\usepackage{color}
\usepackage{soul}
\usepackage[margin=1.5cm,top=2.5cm,bottom=2.8cm,centering]{geometry}
\usepackage[english,francais]{babel}
\usepackage[font=small,labelfont=bf]{caption}
\usepackage[english]{babel}
\usepackage{setspace}
\usepackage{rotating}

 
\begin{document}

\title{\begin{Large} Galilean and relativistic Doppler/aberration effects deduced from spherical and ellipsoidal wavefronts respectively \end{Large}}

\author{Denis Michel \ \href{https://orcid.org/0000-0002-0643-8025}{\includegraphics[scale=0.05]{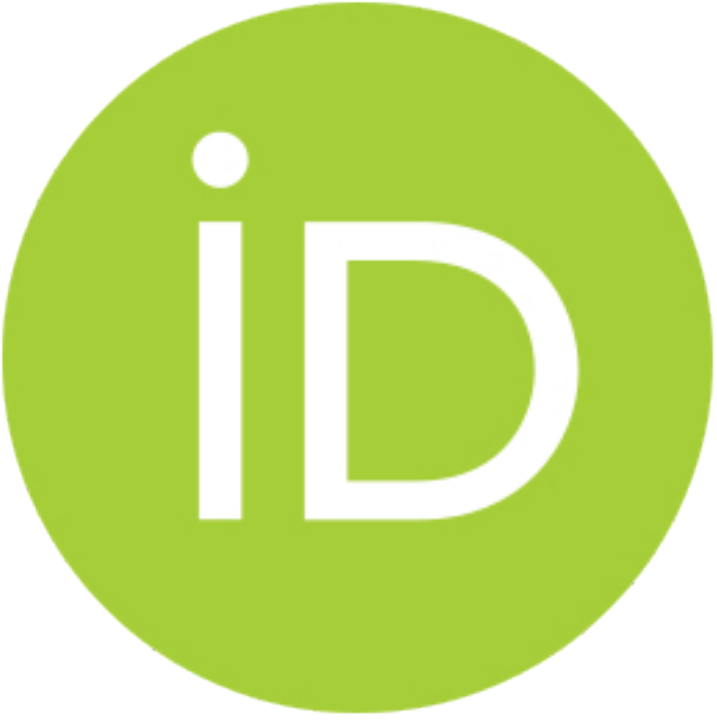}}}
\date{\begin{small} Université de Rennes1, Irset, Rennes, France. E-mail: denis.michel@live.fr \end{small}}
\maketitle
\begin{multicols}{2}
\noindent
\textbf{The diagram showing off-center nested spheres which is traditionally used to illustrate the Doppler effect, is misleading and its trigonometric analysis leads to errors concerning light, because electromagnetic Doppler and aberration effects conform to a wavefront surface that is not a sphere but an ellipsoid stretched along the trajectory of the source. The Cartesian and polar equations of the spherical and ellipsoidal wavefronts are compared here and related to their respective angular Doppler functions. As wavefront surfaces directly link inter-frame coordinate transformations to the aberrations they generate, the simple analysis of their geometry is sufficient to find exact results of special relativity and incidentally to revise the classical aberration formula.\\} 

\noindent
\textbf{Keywords:}\\
\begin{small} Transversal Doppler effect; light aberration; elliptic wavefront; special relativity.\end{small} \\

\noindent
\textbf{Highlights:}\\ \begin{small}
$ \bullet $ Galilean and relativistic Doppler effects are recovered from the geometric analysis of their wavefronts.\\
$ \bullet $ The shape of wavefronts converts coordinate transformations into angular aberrations.\\
$ \bullet $ New reciprocal Galilean aberration and Doppler formulas are established.\\ 

\end{small}
\vspace{-0.5cm}
\section{Introduction}
The scheme widespread in textbooks to explain the Doppler effect represents a series of nested circles, as shown in Fig.1. Assuming that a wave propagates in concentric circles, like rings in water enlarging from a resonator, the spacing between wave crests ($ \lambda $) is the same at all points around a stationary source; but when the source moves, the circles are no longer concentric but shifted backwards, thereby stretching the apparent wavelength behind the source and shortening it in front. But as shown here, these nested circles do not agree with the rules of light aberration and Doppler effects.

\label{fig:profile}
\begin{center}
\includegraphics[width=4.2cm]{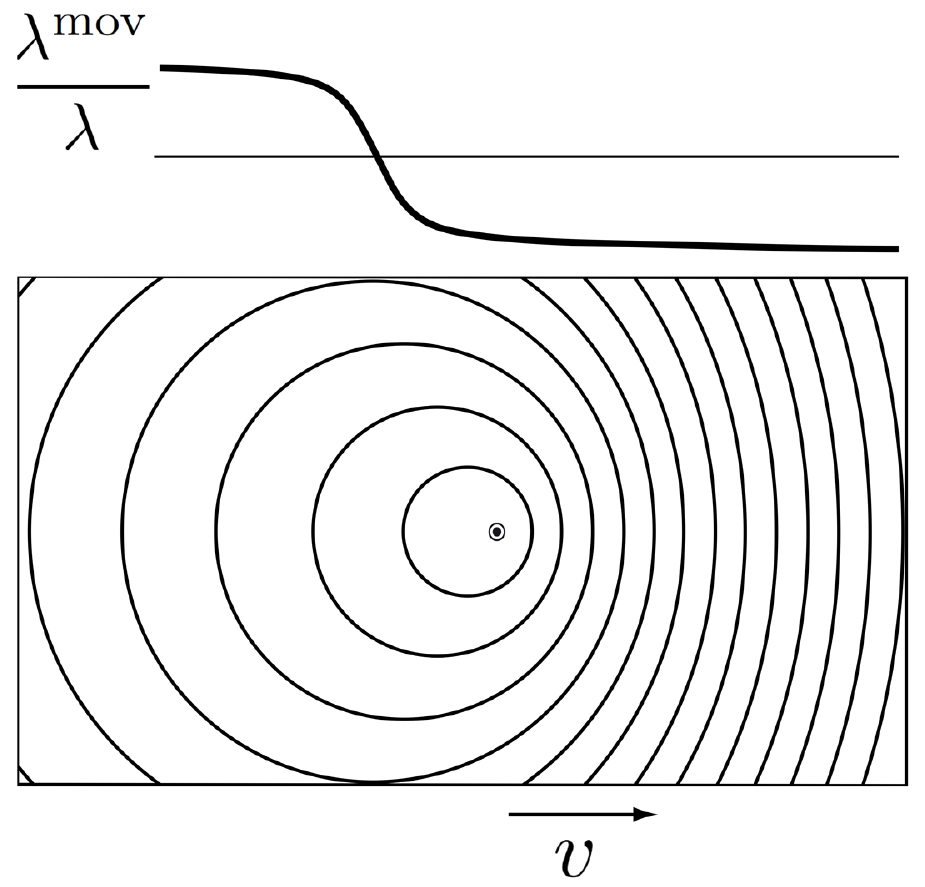}
\end{center}
\begin{small}\textbf{Figure 1}. Evolution of the Doppler effect along the source trajectory schematized for supposedly spherical waves. The Doppler profile for the wavelengths is drawn at the top of the diagram. \\ \end{small} 

\label{fig:SW}
\begin{center}
\includegraphics[width=4.6cm]{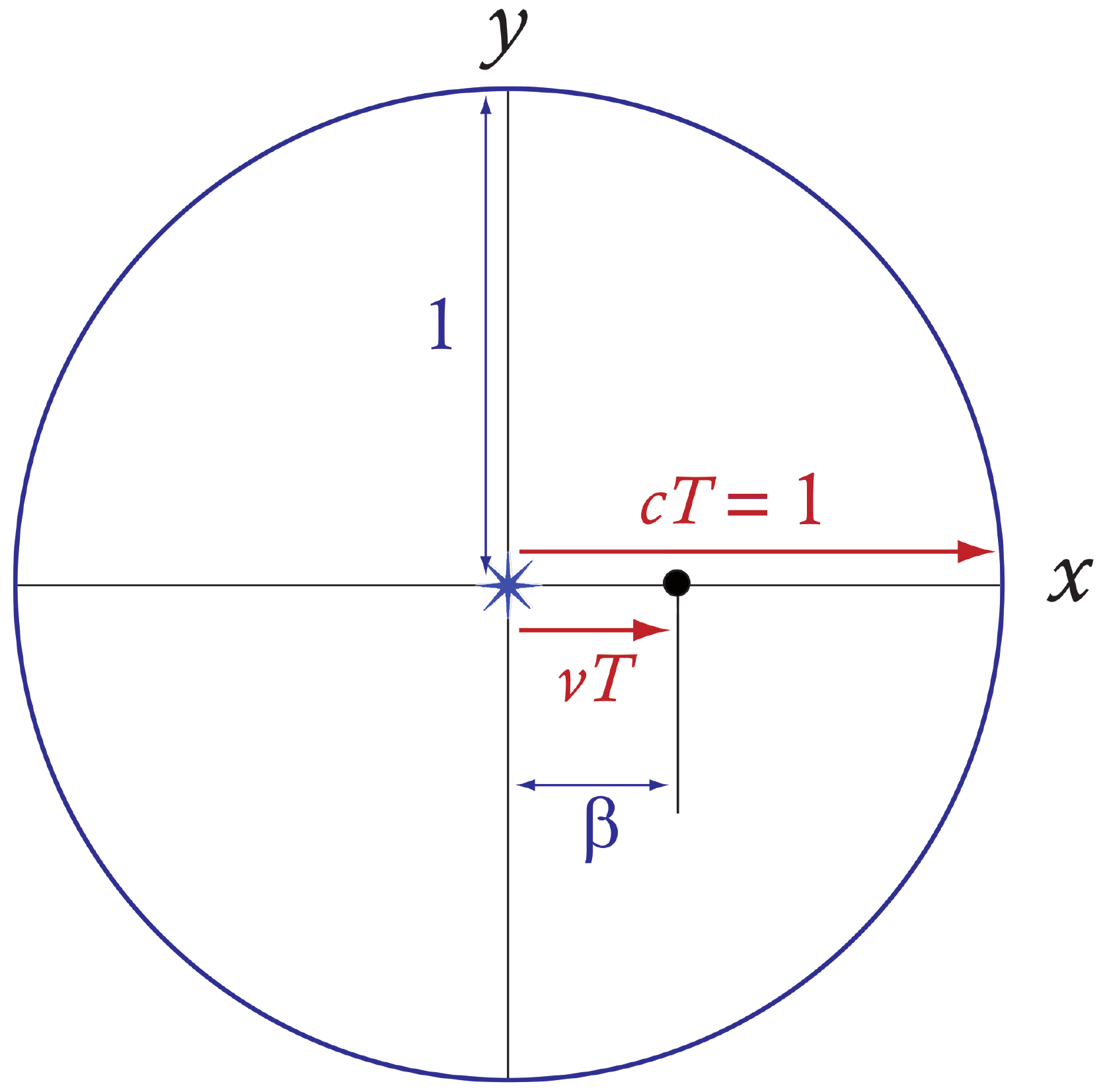}
\end{center}
\begin{small}\textbf{Figure 2}. Analysis of the smallest circle of Fig.1, corresponding to a single period spherical wavefront. The wavefront propagates for a period $ T $ at the same speed $ c $ in all directions around its emission point (central star). During this time, the source has continued to advance (from left to right in this scheme) at speed $ v $, so that the instantaneous image of the system at time $ t=T $ is a circle in which the source is off-center with respect to the emission point in the ratio $ v/c=\beta $. \\ \end{small} 

\section{Spherical wavefronts}

To describe these wavefronts, let us consider the smallest sphere in the diagram of Fig.1, which corresponds to a propagation time of one period $ T $. One period after its emission, the wave will have travelled a distance arbitrarily set to one unit: $ R=cT=1 $, while the source will have only travelled a distance $ vT $ (Fig.2).

\subsection{Equations of spherical wavefronts and their Doppler effect}

The simplicity of the equation of the circle in Fig.2 is related to the simplicity of the Galilean transformations. They read\\

$ x'=x+vt, \ y'=y, \ z'=z $ and $ t'=t $\\

\noindent
Neglecting $ z $, they give the following circle equation

 $$ (x+vt)^{2}+y^{2}=(ct)^2  $$

Assuming that $ c=1 $ and noting the velocity $ v/c = \beta $, for a single unitary period $ t=T=1 $, the Cartesian equation of the offset circle simply becomes

\begin{subequations}  
\begin{equation} (x+\beta)^{2}+y^{2}=1 \label{Eq:cart-classiq} \end{equation}
This equation is convertible in two lines into polar coordinates by noting $ x= \rho \cos \theta $ and $ y= \rho \sin \theta $ (Eq.(\ref{Eq:rho-circ})). Since the radius $ \rho $ corresponds to one wavelength, we have $$ \dfrac{\rho}{R} = \dfrac{cT^{\textup{mov}}}{cT} = \dfrac{\lambda ^{\textup{mov}}}{\lambda} $$ and since for any given angle $ \theta $, the consecutive wavelengths are all identical at constant velocity, $ \rho $ can be considered as the Doppler effect for spherical waves. 
\begin{equation} \rho= \sqrt{1-(\beta \sin \theta)^{2}} - \beta \cos \theta = \dfrac{\lambda ^{\textup{mov}}}{\lambda} \label{Eq:rho-circ} \end{equation} 
\end{subequations} 
\noindent
This formula derived in a different manner by \cite{Klinaku}, has been proposed by this author to be the general formula of Doppler effects, valid as well for light and sound \cite{Klinaku}. It should be noted, however, that this formula denies two major aspects of the Doppler effect of light predicted by special relativity \cite{Einstein}. On the one hand, it does not include the phenomenon of time dilation of periods which specifically stretches the wavelengths seen from a moving source $ \lambda^{\text{mov}} =cT^{\text{mov}} $ with respect to the wavelength at rest. This point could be resolved by a global correction by the Lorentz factor:

\begin{equation}  \dfrac{\lambda ^{\textup{mov}}}{\lambda} = \dfrac{\rho}{\sqrt{1-\beta^{2}}} \end{equation}
\noindent
but on the other hand, this pseudo-relativistic Doppler effect still suffers from another problem: It contradicts the relativistic light aberration. 

\subsection{Aberration of spherical wavefronts }

In the colored right triangle of Fig.3,

\begin{subequations}  
\begin{equation} \cos \theta' = \dfrac{\beta + \rho \cos \theta}{R} \end{equation}
For $ R=1 $ and expressing $ \rho $ as function of the angle $ \theta $ (Eq.(\ref{Eq:rho-circ}))
\begin{equation} \cos \theta' = \cos \theta \sqrt{1-(\beta \sin \theta)^{2}} + \beta \sin^{2} \theta \label{Eq:spher-aber-a}\end{equation}
and conversely
\begin{equation} \cos \theta = \dfrac{\cos \theta' - \beta}{\sqrt{1+\beta^{2}-2\beta \cos \theta' }} \label{Eq:spher-aber-b} \end{equation}
\end{subequations} 

\begin{center}
\includegraphics[width=6.5cm]{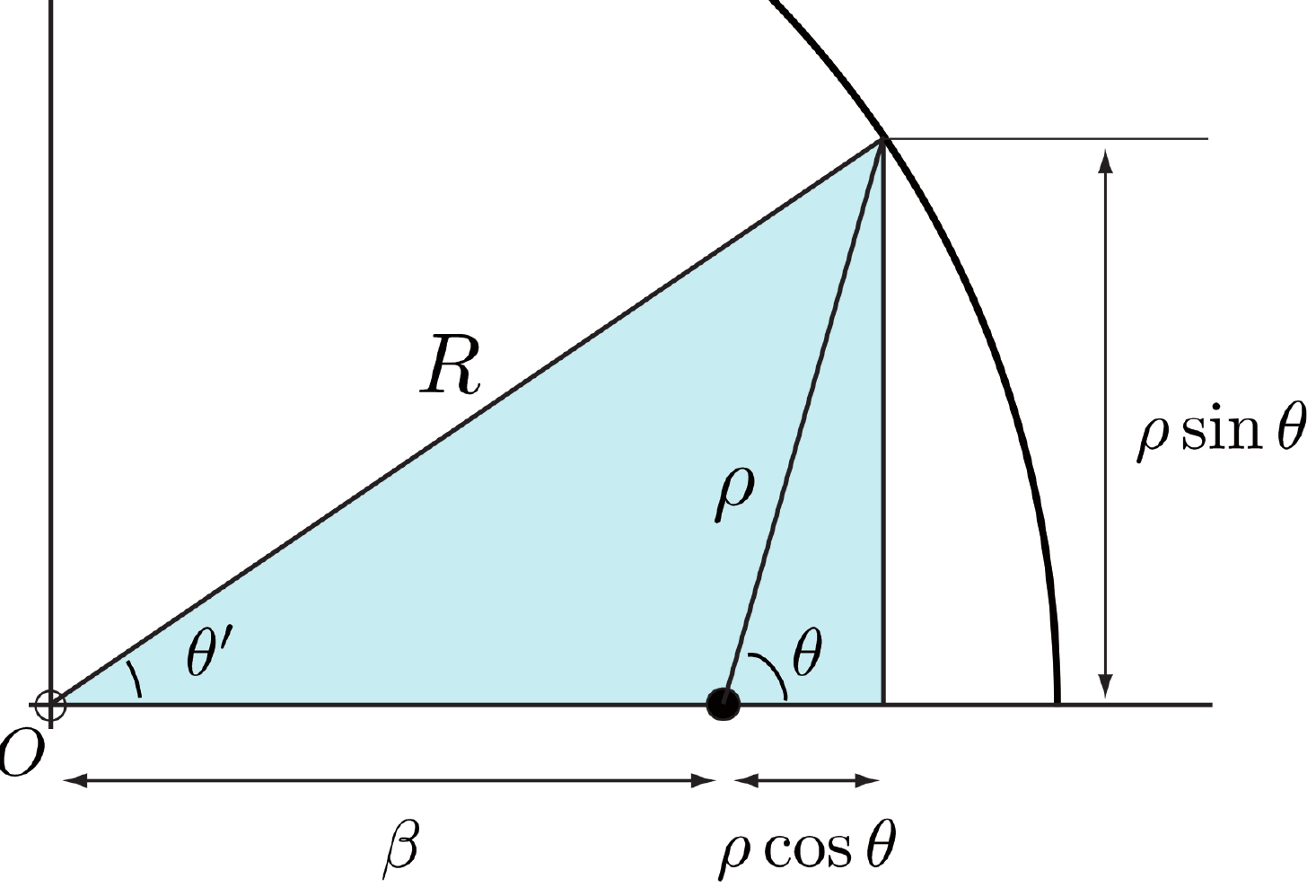} \\
\label{fig:abersphere}
\end{center}
\begin{small} \textbf{Figure 3}. Geometric support for the calculation of the angular aberration generated by a spherical wavefront. \end{small}\\
 
The correspondence pointed out by Einstein between the angle from the source $ \theta = \cos^{-1} (-\beta) $ and the reception angle $ \theta ' = \pi/2 $ \cite{Einstein}, is clearly not verified by these formulas since the Pythagorean theorem as well as Eq.(\ref{Eq:spher-aber-b}) predict that the right angle of reception corresponds to an angle from the source $ \theta = \cos^{-1} \dfrac{-\beta}{ \sqrt{1+\beta^{2}}}$. The tangent function generally used to describe aberration, is also obvious from Fig.3 

\begin{subequations}  
\begin{equation}  \tan \theta' = \dfrac{\rho \sin \theta}{\beta + \rho \cos \theta} \end{equation}
\begin{equation} \tan \theta' =  \dfrac{\sin \theta \left(\sqrt{1-(\beta \sin \theta)^{2}} -\beta \cos \theta \right)}{\cos \theta \sqrt{1-(\beta \sin \theta)^{2}}+ \beta \sin^{2} \theta } \label{Eq:aber-spher} \end{equation}
\end{subequations} 

This equation is not that usually reported in the literature, which ignores the value of $ \rho $, 
$$ \tan \theta' = \dfrac{sin \theta}{\beta + \cos \theta} $$ The latter formula is therefore incomplete and only acceptable for plane sources, like rain falling from the sky or distant sources whose rays appear parallel. Applying the spherical wavefront aberration to the Doppler effect in function of $ \theta $, gives the Doppler effect in function of  $ \theta' $:

\begin{equation}  \dfrac{\lambda ^{\textup{mov}}}{\lambda} =\sqrt{1+\beta^{2}-2 \beta \cos \theta'} \label{Eq:dop-circ} \end{equation}

which could have been directly anticipated by applying Al-Kashi's formula to the angle $ \theta' $ of Fig.3. For $ R=1 $,

\begin{equation} \rho^{2} =1+\beta^{2}-2 \beta \cos \theta' \label{eq:alkashi} \end{equation}

The reciprocal Doppler functions of $ \theta $ (Eq.(\ref{Eq:rho-circ})) and $ \theta '$ (Eq.(\ref{Eq:dop-circ})) are compared on Fig.4, in which some remarkable angles and Doppler values are mentioned to serve as points of comparison with the Doppler effect of light discussed later. Fig.4 also shows that an overlooked property of the Doppler effect of spherical waves is that its cancellation ($ \lambda ^{\textup{mov}}/\lambda =1  $) is obtained neither for $ \theta =\pi/2 $ nor for $ \theta' =\pi/2 $.

\label{fig:SWD}
\begin{center}
\includegraphics[width=9cm]{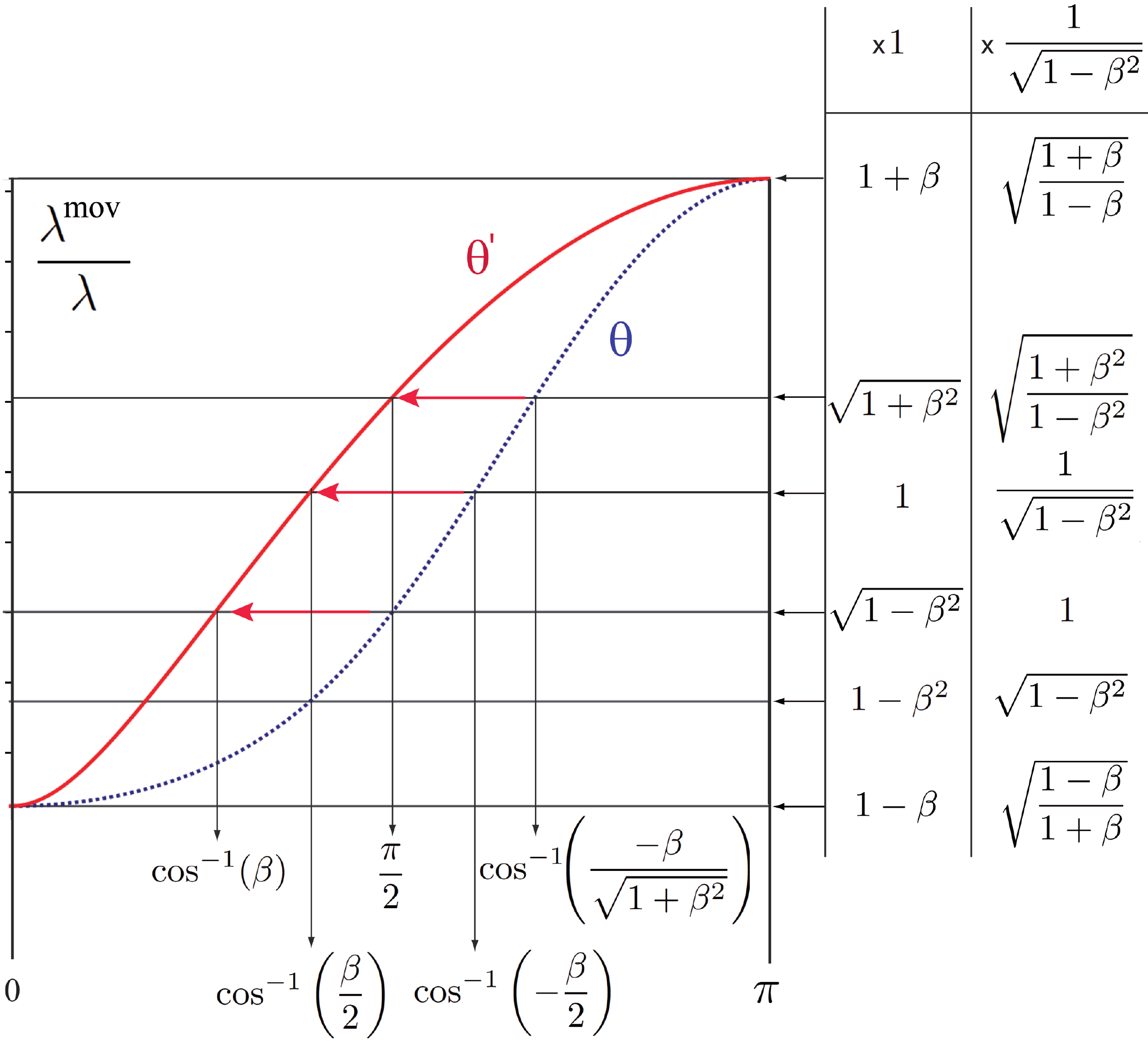}
\end{center}
\begin{small}\textbf{Figure 4}. Doppler effect of spherical wavefronts as a function of the angle $ \theta $ (blue dashed curve) and $ \theta' $ (red plain line). Scheme drawn for a source velocity of two thirds of the wave velocity. The correspondence between the angles $ \theta $ and $ \theta' $ is based on the specific aberration relations of spherical wavefronts calculated here. Some remarkable Doppler effects are mentioned on the right, those indicated x1 are the direct results of Eq.(\ref{Eq:rho-circ}) and Eq.(\ref{Eq:dop-circ}). In the right column, these results are multiplied by the Lorentz factor, only for comparison purposes because this multiplication has no physical basis. \end{small} 

\section{Ellipsoidal wavefronts}

The simple correspondence between the angles $ \theta = \cos^{-1} (-\beta) $ and $ \theta ' = \pi/2 $ is enough to predict a lateral flattening of the wavefront by $ \sqrt{1-\beta^{2}} $ with respect to its dimension collinear with the source translation. Indeed, the radius $ b $ of the bubble orthogonal to the trajectory must verify 

\begin{equation} b = \beta \ \tan \ (\cos^{-1} -\beta)= -\sqrt{1-\beta^{2}} \end{equation}

This result shows that the light wave bubble is not spherical but deformed so that the radius orthogonal to the trajectory of the source is flattened by $ \sqrt{1-\beta^{2}} $ compared to the radius along the trajectory, which explains why the spherical wavefront cannot be consistent with the aberration of light. It is an ellipse of which a focus is precisely occupied by the source. The hypothesis of an ellipsoidal wavefront has been episodically put forward \cite{Poincare1905,Moreau,Pierseaux}, sometimes critically \cite{Walter}, but generally with little success. Yet the relevance of a theory should be appreciated through the accuracy of its predictions and precisely, the ellipsoid wavefront gives the exact relativistic Doppler and aberration results. Poincaré explained that the elliptic shape of the light wavefront results paradoxically from the contraction of the lengths which makes that a material body, of  spherical shape when it is at rest, will become an ellipsoid of rotation flattened in the direction of its motion, but that the observer will always see it spherical because he undergoes himself an analogous deformation, as well as all the objects which are used as reference points and tools of measurement. In contrast, the wave surface of light that have remained rigorously spherical, is expected to become an ellipsoid elongated in the direction of the source's motion \cite{Poincare1918}. In the section 8 dealing the radiation pressure of his celebrated article of 1905, Einstein also explained that a spherical wave, viewed in the moving system, is an ellipsoid surface \cite{Einstein}. To define the simplest Cartesian equation for such an ellipsoid, in addition to applying an offset $ +\beta $ to the $ x$ axis as done previously for the spherical bubble (Eq.(\ref{Eq:cart-classiq})), we will apply specifically to this coordinate the Lorentz - Fitz-Gerald contraction factor $ \sqrt{1-\beta^{2}} $. This contraction derives naturally from the famous relativistic transformations which give a particular role to the $ x $ coordinate. Lorentz established transformations which do not alter the equations of the electromagnetic field \cite{Lorentz} and Poincaré wrote them so as to make them a mathematical group. Considering that the limit velocity is $ c=1 $ and ignoring the units, $ v= v/c = \epsilon $ (replaced here for clarity by the conventional notation $ \beta $), he wrote \cite{Poincare1905}, \\

$ x'=\dfrac{x+\beta t}{\sqrt{1-\beta^{2}}}, \ y'=y, \ z'=z $ and $ t'=\dfrac{t+\beta x}{\sqrt{1-\beta^{2}}} $\\

\noindent
Substituting $ t $ in the first transformation by its value given in the last one, gives
 
\begin{equation} x'=x\sqrt{1-\beta^{2}} + \beta t' \label{Eq:x'}\end{equation}

This modifications of $ x' $ gives for a unit period, the wavefront surface equation

\begin{subequations}
\begin{equation} (x\sqrt{1-\beta^{2}}+\beta)^{2}+y^{2}+z^{2}=1 \label{Eq:cart-relat} \end{equation} 

which describes an ellipsoid with symmetry of rotation around the $ x $ direction, so that we can neglect $ z $ without loss of generality,

\begin{equation} (x\sqrt{1-\beta^{2}}+\beta)^{2}+y^{2}=1 \label{Eq:cart-relat} \end{equation} 
\end{subequations}
Eq.(\ref{Eq:cart-relat}) is an the ellipse 
\begin{itemize}
\item of center: $ (x_{0},y_{0}) = (- \beta, 0) $
\item of foci (on $ x $): $ x_{0}-\dfrac{\beta}{\sqrt{1-\beta^{2}}} $ and $ x_{0}+\dfrac{\beta}{\sqrt{1-\beta^{2}}} $
\item of eccentricity (on $ x $): $ \beta $
\item of semiminor axis length (on $ y $): 1
\item of semimajor axis length (on $ x $): $ 1/\sqrt{1-\beta^{2}} $
\item of area enclosed: $ \pi/\sqrt{1-\beta^{2}} $
\\
\end{itemize}

Replacing $ x $ and $ y $ in the Cartesian equation with $ r \cos \theta $ and $ r \sin \theta $ and then solving for the unknown $ r $, yields its polar equation. The powers of $ r $ are segregated to form the quadratic equation\\

$ r^2(1-\beta^{2} \cos^{2} \theta) + r \ (2 \beta (1-\beta^{2}) \cos \theta) - (1-\beta^2) =0 $ \\

\noindent
whose solution remarkably simplifies

\begin{equation} r = \dfrac{\sqrt{1-\beta^{2}}}{1+\beta \cos \theta } = \dfrac{\lambda ^{\textup{mov}}}{\lambda} \label{Eq:r-relat} \end{equation}

The Doppler equation of Einstein expressed as a function of the angle $ \theta $ \cite{Einstein} is identical to the polar equation of the ellipse. The elliptical and circular wavefronts are compared in Fig.5 for different values of $ \beta $.

\label{fig:ellipsphere}
\begin{center}
\includegraphics[width=9cm]{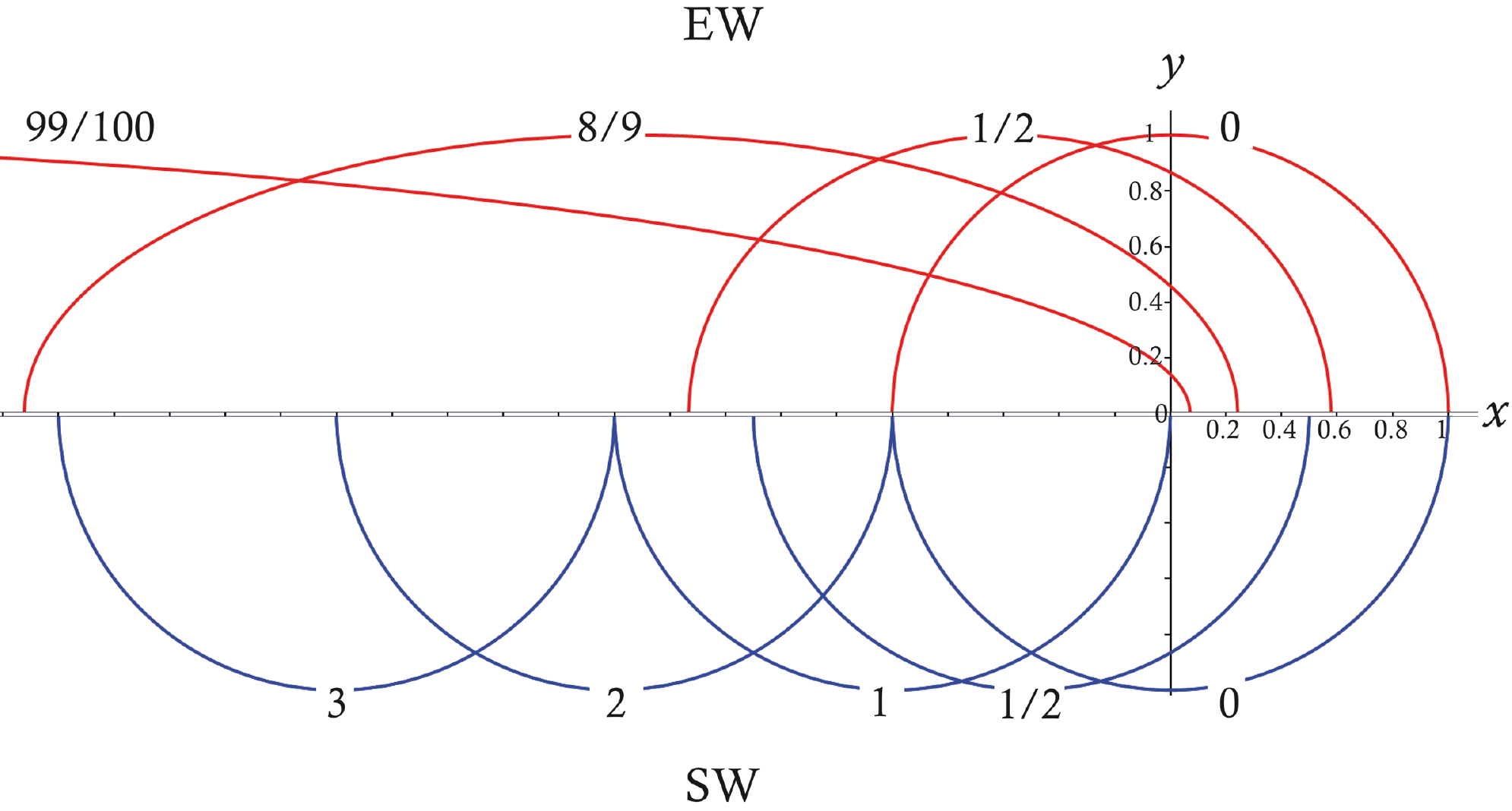}
\end{center}
\begin{small}\textbf{Figure 5}. Comparison of single period elliptical (EW) and spherical (SW) wavefronts for different values of $ \beta $. $ \beta$ can exceed 1 for the circle. Beyond this value, the decentering of the circles is accompanied by their separation. By contrast, $ \beta =1 $ is obviously a limit value for the ellipse, at which the orthogonal dimension $ y=1 $ is reached at an infinite distance backwards. \\ \end{small} 

\section{Comparison of angles between circular and elliptical wavefronts}

Compared to the sphere, the relativistic bubble is wider on the $ x $ axis by $ 1/\sqrt{1-\beta^{2}} $, which makes the geometric comparison difficult. To make their diameters coincide on $ x $ while preserving the proportions of the ellipse, let us just contract the radius orthogonal to the trajectory, from 1 to $ \sqrt{1-\beta^{2}} $. For a source moving from left to right, the Cartesian equation of this simplified ellipsoid is

\begin{equation} (x+\beta)^{2}+\dfrac{y^{2}}{1-\beta^{2}}+\dfrac{z^{2}}{1-\beta^{2}}=1 \label{Eq:cart-ellipse2} \end{equation} 

\begin{itemize}
\item of center: $ (x,y,z) = (- \beta, 0, 0) $
\item of eccentricity (on $ x $): $ \beta $
\item of semimajor axis length (on $ x $): 1
\item of semiminor axis length (on $ y $ and $ z $): $ \sqrt{1-\beta^{2}} $
\item of foci (on $ x $): $ -2\beta $ and $ 0 $\\
\end{itemize}

Restricting ourselves to the two-dimensional ellipse, replacing once again $ x $ and $ y $ in the Cartesian equation by $ \rho \cos \theta $ and $ \rho \sin \theta $ and then solving for the unknown $ \rho $, gives the corresponding polar equation. Let us first multiply the two sides of Eq.(\ref{Eq:cart-ellipse2}) by $ 1-\beta^2 $:\\

\noindent
$ (1-\beta^{2})(\rho^2 \cos^{2} \theta +2 \beta \rho \cos \theta +\beta^{2})+ \rho^2 \sin^{2} \theta =1-\beta^2 $\\

\noindent
from which\\

\noindent
$ \rho^2(1-\beta^{2} \cos^{2} \theta) + \rho \ (2 \beta (1-\beta^{2}) \cos \theta) - (1-\beta^2)^2 =0 $\\

\noindent
whose solution is

\begin{equation} \begin{split}\rho & = \dfrac{(1-\beta^{2})\left(-\beta \cos \theta +\sqrt{\beta^{2}\cos^{2} \theta+(1-\beta^{2} \cos^{2} \theta)}\right)}{1-\beta^{2} \cos^{2} \theta} \\& = \dfrac{(1-\beta^{2})(1-\beta \cos \theta)}{(1-\beta \cos \theta)(1+\beta \cos \theta)}\\& = \dfrac{1-\beta^{2}}{1+\beta \cos \theta } \end{split} \label{Eq:rho-ellip} \end{equation}

\noindent
which can be considered as the primary Doppler effect. The secondary effect of dilation by the Lorentz factor of the periods $ T ^{\textup{mov}} $, gives the complete Doppler equation.

\begin{equation} \dfrac{\lambda ^{\textup{mov}}}{\lambda} = \dfrac{cT^{\textup{mov}}}{cT}= \dfrac{\rho}{\sqrt{1-\beta^{2}}} = \dfrac{\sqrt{1-\beta^{2}}}{1+\beta \cos \theta } \label{Eq:rho-relat}  \end{equation}

\begin{center}
\includegraphics[width=8cm]{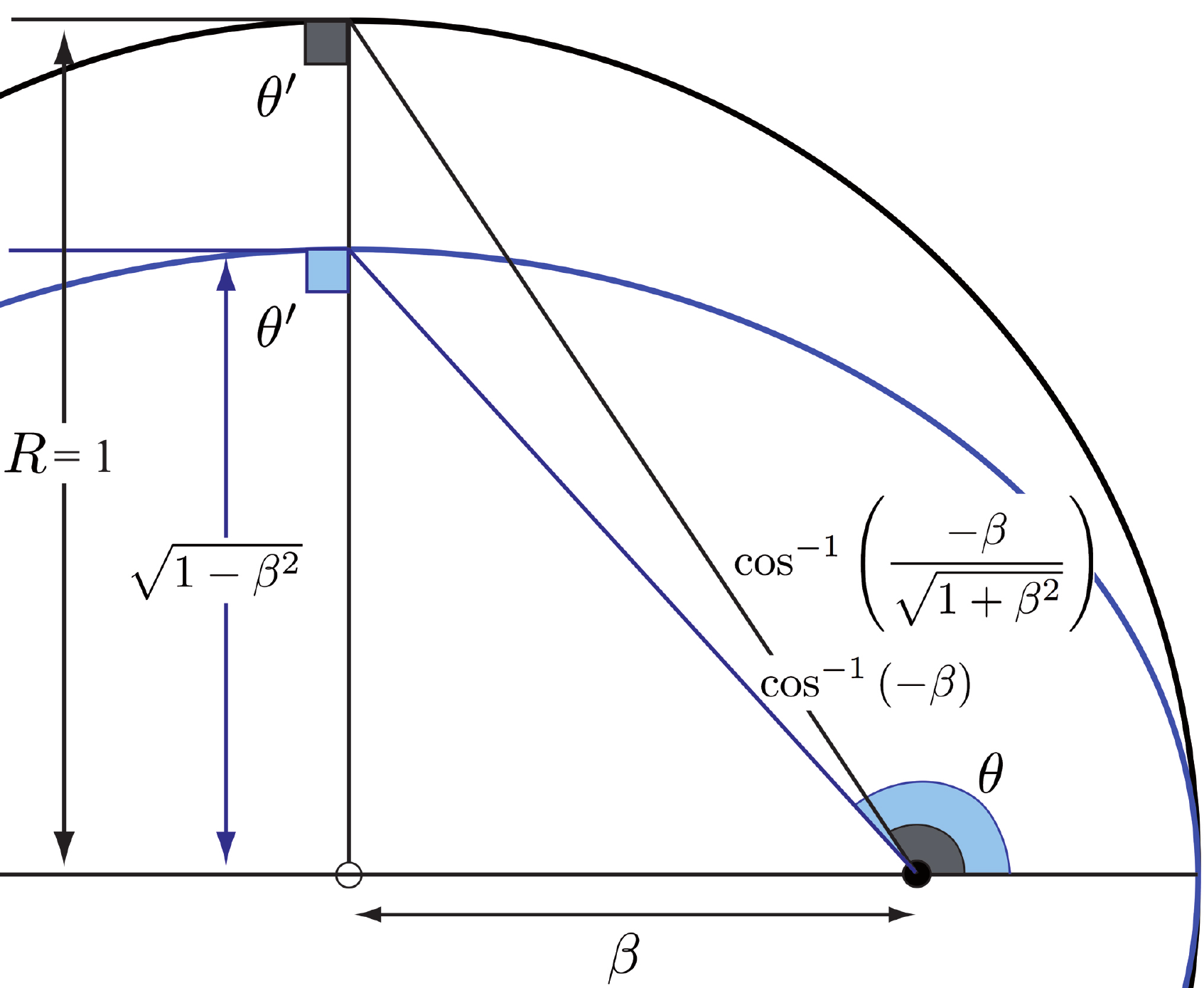} \\
\label{fig:spherellipse}
\end{center}
\begin{small} \textbf{Figure 6}. Comparison of the angles $ \theta $ giving a reception angle of $ \theta'=\pi/2 $, between spherical and elliptical bubbles. \end{small}\\

\noindent
As it is not directional, the relativistic dilation transforms the ellipse in a homothetic way without altering its characteristics and angles. So once adjusted, it is possible to easily compare the angles $ \theta $ giving a reception at right angle to the trajectory (Fig.6). The spherical and elliptical equations agree with simple predictions of the Pythagorean theorem. $ \theta'=\pi/2 $ corresponds for the sphere to $ \theta= \cos^{-1} (-\beta/ \sqrt{1+\beta^{2}}) $ and for the ellipse to $ \theta= \cos^{-1} (-\beta )$. For example, for a source velocity of $ 2 \times 10^{8} $ m/s ($ \beta $= 2/3), $\theta$ is 2.3 rad for the ellipse and 2.1 rad for the sphere. 

\section{Geometrical analysis of the ellipsoidal wavefront} 

\subsection{The relativistic aberration}

The geometric determination of aberration is a little more subtle for the elliptical wavefront than for the spherical one, since except for the particular case of $ \theta = \cos^{-1} \beta $, the intersection between the focal radius $ \rho $ and the perimeter of the ellipse does not correspond to a point tangential to the radius emanating from the center of the ellipse $ O$. It is therefore necessary to consider the fixed reference frame of this center, which is a point of emission and as such is an event, devoid of velocity and duration from which the wave front propagates with spherical symmetry (Fig.7).

\begin{center}
\includegraphics[width=8cm]{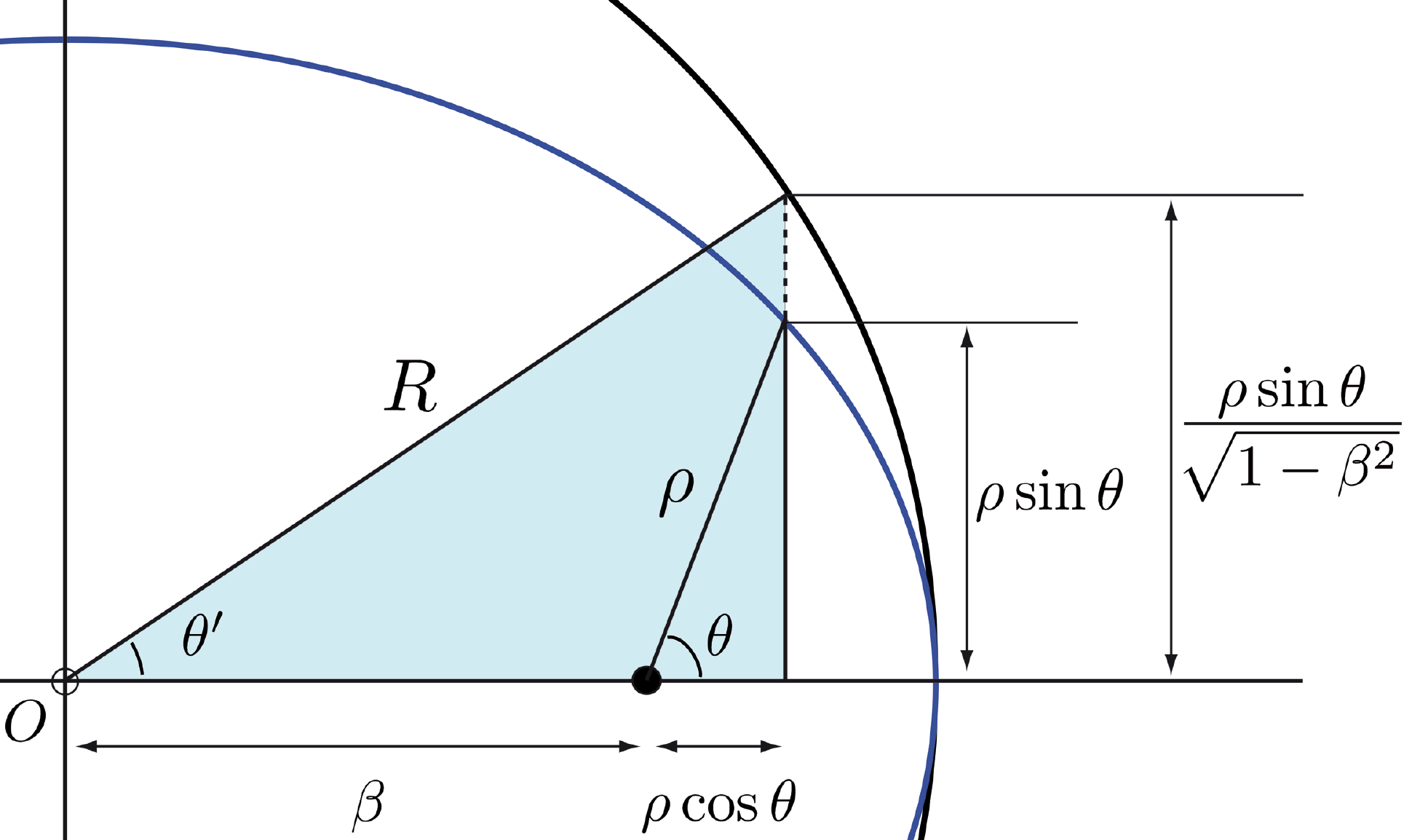} \\
\label{fig:aberellipse}
\end{center}
\begin{small} \textbf{Figure 7}. Geometric support for the simplified calculation of relativistic aberration. \end{small}\\

The segments $ \beta $ and $ \rho \cos \theta $ do not undergo the flattening and remain unchanged in this representation. We have therefore
\begin{subequations}  
\begin{equation}  \cos \theta' =  \dfrac{\beta + \rho \cos \theta}{R}\end{equation}
As $ R $ is the large radius fixed at 1 and the value of $ \rho $ is given in Eq.(\ref{Eq:rho-ellip}), this equation becomes

\begin{equation} \cos \theta' = \dfrac{\cos \theta + \beta}{1+\beta \cos \theta}  \label{Eq:aber-cos} \end{equation} 
whose reciprocal form is as elegant.
\begin{equation} \cos \theta = \dfrac{\cos \theta' - \beta}{1-\beta \cos \theta'}  \label{Eq:aber-cos2} \end{equation} 
\end{subequations}

By contrast, the segment $ \rho \sin \theta $ where $ \rho $ takes into account the flattening by $ \sqrt{1-\beta^{2}} $, must be corrected by the inverse factor to be prolonged to the circle.

\begin{equation} \begin{split} \sin \theta' & =   \dfrac{\rho \sin \theta }{\sqrt{1-\beta^{2}} R} \\ & = \dfrac{(1-\beta^{2})\sin \theta}{(1+\beta \cos \theta)\sqrt{1-\beta^{2}}} \\ & = \dfrac{\sqrt{1-\beta^{2}}\sin \theta }{1+\beta \cos \theta}, \end{split} \label{Eq:aber-sin} \end{equation}
\noindent
in agreement with the correspondence between the sine and cosine 

$$ \sin \theta' = \sin \left(\cos^{-1}\left( \dfrac{\cos \theta + \beta}{1+\beta \cos \theta}\right)\right) = \dfrac{\sqrt{1-\beta^{2}}\sin \theta }{1+\beta \cos \theta} $$
\noindent
Eq.(\ref{Eq:aber-cos}) and Eq.(\ref{Eq:aber-sin}) give the usual relation
\begin{equation} \tan \theta'= \dfrac{\sin \theta'}{\cos \theta'}= \dfrac{\sin \theta \sqrt{1-\beta^{2}}}{\beta + \cos \theta}  \end{equation}

Strikingly, in spite of a slightly more elaborate Cartesian equation, the elliptic aberration formula turns out to be much simpler and more elegant than that of its \underline{true} classical counterpart given by Eq.(\ref{Eq:aber-spher}). Moreover, once the ellipse hypothesis is introduced, these results are obtained in a simplistic manner without recourse to the tools of special relativity. In particular, conventional additions are used instead of the velocity composition theorem. As a reminder, a more standard relativistic approach would be

\begin{equation} \begin{split} c \cos \theta'  &= v \oplus c \cos \theta  \\ & = \dfrac{v + c \cos \theta}{1+\dfrac{v c \cos \theta}{c^{2}}}   \end{split}  \end{equation}
which gives upon division by $ c $, the Eq.(\ref{Eq:aber-cos}). Fig.8 shows the differences, perceptible only at significant speed, between the relativistic and classical aberration formulas recalculated here. In both formulas, except for 0 and $ \pi $, $ \theta' $ is always smaller than $ \theta $. The only pairs ($ \theta, \theta' $) common to both formulas are $ (0 , \ 0)$, $ (\pi/2 , \cos^{-1} \beta )$ and $ (\pi, \pi)$. For $ \theta < \pi/2 $, the relativistic $ \theta' $ is greater than the Galilean $ \theta' $, whereas for $ \theta > \pi/2 $, the relativistic $ \theta' $ is less than the Galilean $ \theta' $.

\subsection{The relativistic Doppler effect}

The incorporation of Eq.(\ref{Eq:aber-cos2}) in Eq.(\ref{Eq:rho-relat}) allows to express the Doppler formula as a function of $ \theta' $

\begin{equation} \dfrac{\lambda ^{\textup{mov}}}{\lambda} =  \dfrac{\sqrt{1-\beta^{2}}}{1+\beta \cos \theta }= \dfrac{1-\beta \cos \theta' }{\sqrt{1-\beta^{2}}}\end{equation}

As Doppler effects are rarely compared in their entirety from 0 to $ \pi $ for the angles $ \theta $ and $ \theta' $, these two functions are compared in Fig.9. Note that for $ \theta $ as well as for $ \theta' $, the cancellation of the Doppler effect ($ \lambda ^{\textup{mov}}/\lambda =1 $) depends on the speed of the source. None of the angles giving this null Doppler effect is the solution reported by \cite{Nolte}: $ \cos^{-1} (\beta/2) $, which results again from the misuse of spherical trigonometry (see Fig.4) in a situation where it does not apply. 

\label{fig:comparaber}
\begin{center}
\includegraphics[width=8.2cm]{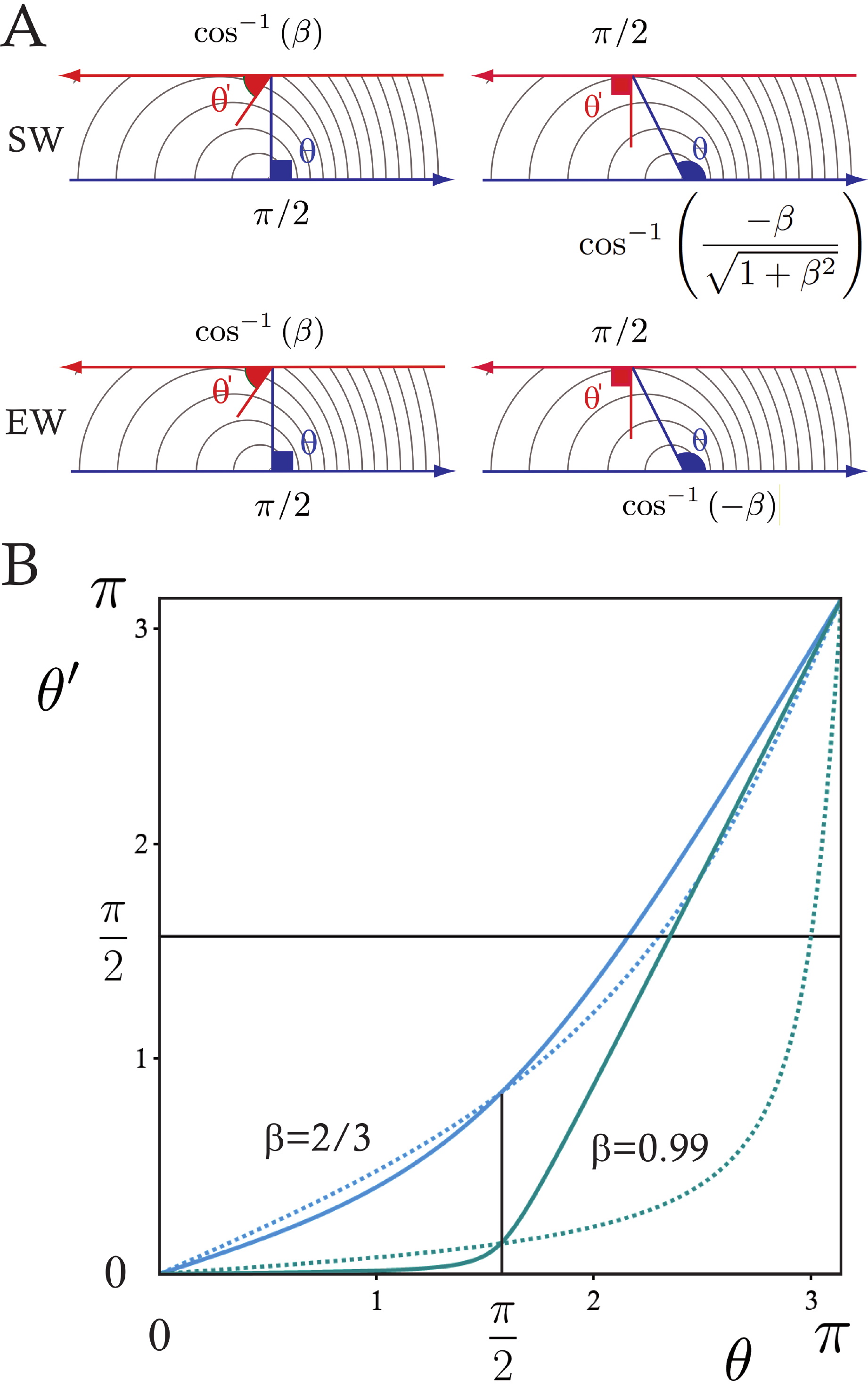} \\
\end{center}
\begin{small} \textbf{Figure 8}. Compared aberration effects of the spherical (SW) and elliptical (EW) wavefronts. (\textbf{A}) Explanatory diagrams of the angular shift related to the aberration phenomenon for two remarkable points involving the angle $ \pi/2 $. These schemes are drawn for $ \beta =1/3 $. The lower arrows going from left to right represent the trajectory of the source and the upper arrows represent the relative trajectory of the detector. (\textbf{B}) Complete profiles of the elliptical (dotted curves) and spherical (plain curves) aberration curves, for $ \beta =2/3 $ (upper blue curves) and $ \beta =99/100 $ (lower green curves). For any value of $ \beta $ apart from 0 and $ \pi $, the only common point of the two curves is $ (\theta, \ \theta') =(\pi/2, \ \cos^{-1} \beta )$ \end{small}\\

\section{Discussion}

The exhaustive representation of Doppler curves in Fig.9 clearly shows that the identity between the transverse light Doppler effect and the Lorentz dilation factor, expected by Einstein as the confirmation of his theory \cite{Einstein2}, is received, but not emitted, at right angle to the trajectory, in accordance with the fact that time dilation is a phenomenon taking place between different reference frames. It is therefore crucial, in an experimental device aimed at demonstrating it, not to canalize the emitted light at right angle to the trajectory, as in the device of \cite{Hasselkamp}. The comparison of the Galilean and relativistic Doppler effects is also very instructive. It is often explained in the literature that the relativistic Doppler effect is the classical Doppler effect corrected by the Lorentz factor. The comparison of Fig.4 and Fig.9 shows that this is not true. The simple multiplication of the Galilean Doppler values by the Lorentz factor (right column of Fig.4) may seem satisfactory when considering only the collinear effects (angles 0 and $ \pi $), but does not recapitulate the relativistic Doppler effect for all the other points (Fig.9), because the main difference between these Doppler effects is angular and precisely related to the spherical vs ellipsoidal nature of their wavefronts. 

\label{fig:TRD}
\begin{center}
\includegraphics[width=9cm]{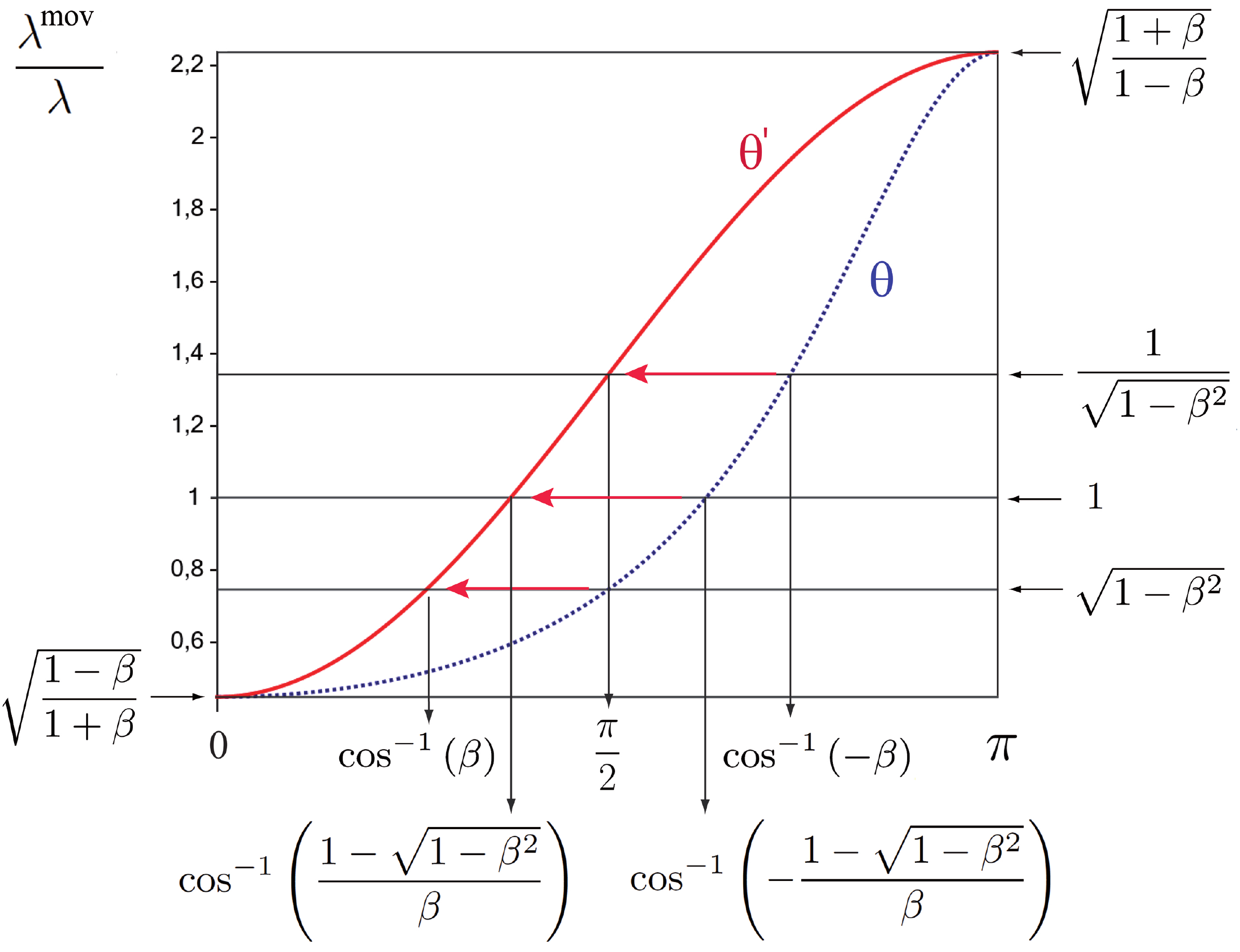} \\
\end{center}
\begin{small} \textbf{Figure 9}. Relativistic Doppler effect expressed as a function of $ \theta $ (blue dotted curve) and $ \theta' $ (red continuous curve) for a source velocity of two thirds of light velocity. \end{small}\\

Poincaré proposed that the shortening of distances in the direction of translation also affects our measuring devices as well as our standard meters, so that it goes unnoticed. But if lengths are no longer measured with standard meters but by the travel time of light, then this shortening does not apply, which accounts for both: (1) the elliptical nature of the light propagation bubble and (2) the negative result of the Michelson-Morley type experiments, interpreted as the consequence of an exact compensation between the lower speed of light traveling against the ether wind and the contraction of the length to be traveled. The elliptical shape of the relativistic wavefront was deduced from special relativity by Poincaré, and conversely as shown here, its trigonometric analysis confirms other results of special relativity such as the relativistic aberration, relativistic beaming and relativistic Doppler effect. Poincaré's ellipse is therefore an integral part of this theory and as such presents a theoretical interest and has pedagogical virtues. Taking the ellipse into consideration allows to recover essential results without need for sophisticated mathematical tools, and most importantly, avoids errors. Doppler studies ignoring the ellipse and built on circular diagrams, unintentionally deny Lorentz transformations and consequently yield incorrect results. The angular relations which are not obvious at first sight in the coordinate transformations, are precisely introduced by the shapes of the resulting wavefronts.

\end{multicols}

\vspace{1cm}

\section*{Graphical Abstract:}

\begin{center}
\includegraphics[width=18cm]{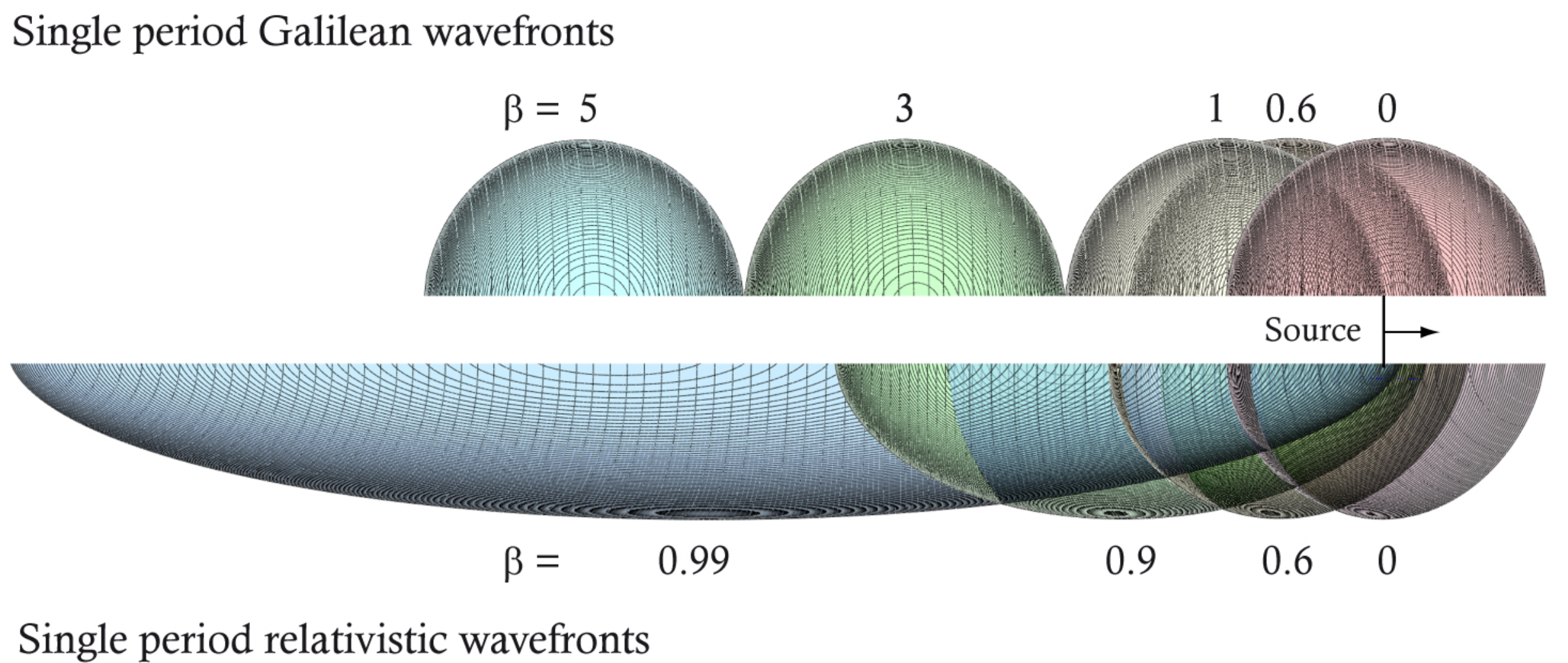} \\
\end{center}

\end{document}